\documentstyle[aps,12pt,preprint]{revtex}

\newcommand{\sect}[1]{\setcounter{equation}{0}\section{#1}}

\def\tg{\tilde g }

\begin{document}

\begin{flushright}
{UT-KOMABA-95/2}\\
{\sl January  1995}\\
\end{flushright}

\begin{center}
{\large\bf
Matrix model approach to the flux lattice melting \\
in $2D$ superconductors}

\vskip 10mm
\vspace{6pt}
{\sc A. Fujita and S. Hikami}

\vskip 10mm

{\sl Department of Pure and Applied Sciences,\\
University of Tokyo, Meguro-ku, Komaba, Tokyo 153, Japan\\}
\end{center}

\vspace{24pt}

\begin{abstract}

We investigate a gauged matrix model in the large $N$ limit which is closely
related to the superconductor fluctuation and the flux lattice melting in two
dimensions.   With the use of saddle point method the free energy is expanded
up to eighth order for the coupling constant $g$.  In the case that the
coefficient of quadratic term of the Ginzburg-Landau matrix model is negative,
a critical point $g=g_c$ is obtained in the large $N$ limit and the relation
between  this phase transition and the 2D flux lattice melting transition is
discussed.

\end{abstract}
\newpage

\sect{Introduction}

In the study of the fluctuation of superconductors and the melting transition
of the Abrikosov flux lattice in a strong magnetic field, Ginzburg-Landau (GL)
model is especially useful with  the lowest Landau level approximation.   This
approximation is valid in the region very near to $H_{c2}$ and neglects all the
contribution from upper Landau levels except the lowest one,  and reduces the
effective dimension of the system by two due to the Landau quantization
perpendicular to the  magnetic field axis.

The perturbational studies \cite{large}-\cite{hu2} for GL free energy up to
high orders are unable to find a flux lattice melting transition in two
dimensions.   However, the comparison of the low temperature expansion to the
high  temperature expansion for the free energy  gives the estimation of the
melting transition point\cite{HFL}.   On the other hand, the numerical studies
such as Monte Carlo simulations show that there exists the first order
transition in a finite temperature well below the mean field superconducting
transition point $H_{c2}$\cite{kato}-\cite{sasik}.   Although both analytical
and numerical calculations agree quite well in derivation of the statistical
amounts, the nature of the melting transition  of flux lattice is not yet fully
understood \cite{tesa}-\cite{oneil}.   The first order transition has  been
suggested by the renormalization group analysis\cite{BNT}, but two dimensions
are  far from the valid region around six dimensions.

In this study we generalize the GL Hamiltonian to a gauged matrix model in
which
the order parameter is expressed by a complex $N\times N$ matrix\cite{hikami1}.
The matrix model in the large $N$ limit attracts recently much theoretical
interest since it has a close relation to the string field theory
\cite{brezin4}
and has been studied in many fields such as $2d$ quantum gravity coupled to
matter field\cite{hikami2}, mesoscopic fluctuations\cite{brezin2} and electron
correlations\cite{okiji}.   In several matrix models the exact  solutions have
been obtained\cite{brezin1} and these solutions give the clues to the
perturbative analysis of other unsolved matrix models.  It is known that the
matrix model has a phase transition in the  large $N$ limit when the
coefficient
of the quadratic term is negative or in the lattice gauge theory\cite{gross}-
\cite{sima}.  This may be true for our gauged matrix model and we are
interested
in how this phase transition is related to the superconducting flux lattice
melting. Usual Ginzburg-Landau model corresponds to $N=1$ case, but we
generalize the order parameter to  $N\times N$ complex matrix and take the
large
$N$ limit.  This large $N$ limit should be distinguished from the large $N$
case
of the $N$-vector model\cite{affl,HF}, which becomes equivalent to the Hartree-
Fock approximation and has no phase transition in two dimensions.

In the usual GL model, the perturbation series becomes asymptotic expansions
and
one needs Borel summation for such divergent series.  In the large $N$ limit of
GL matrix model, the perturbation series about $g$ is convergent, and the
precise analysis of the free energy becomes possible. Thus it is natural to
consider the solution of the GL matrix model in the large $N$ limit as a first
approximation of the $2d$ superconductor phase transition.  We find indeed  a
phase transition point $g_c /\alpha_B^2 = 0.07$ for the large $N$ limit of the
GL matrix model and this point corresponds to $y=-2.7$  in term of the reduced
relative temperature.  In the previous paper\cite{HFL},  we evaluated the flux
lattice melting temperature for $2d$ superconductors ($N=1$) as $y=-10$ by the
analysis of the perturbation series of the usual GL free energy.   Although the
phase transition for $N=\infty$ occurs in considerably higher temperature than
the usual GL model of $N=1$, we consider this difference may result from  the
first approximation for $2d$ superconductors.   The improvement of the
evaluation of this phase transition point in the gauged matrix  model is
suggested by taking the higer order terms in $1/N^2$ expansion.

This paper is organized as follows :  In section II, a gauged matrix model with
the interaction of Ginzburg-Landau type ${g\over N} {\rm Tr} (M^\ast M)^2$ is
introduced in the lowest Landau level  approximation and the free energy is
obtained as the power series of coupling constant $g$ using the saddle point
method.   In section III, the critical point of the free energy is obtained for
the case the coefficient of ${\rm Tr} M^\ast M$ term is negative.  This phase
transition is  investigated in nature and in section IV the relation  to the
melting transition of flux lattice is discussed.

\sect{Gauged matrix model in two dimensions}

In the Ginzburg-Landau model, the order parameter $\phi$ is a complex field
and can be expanded with the Landau levels.  If the field is near the critical
point $H_{c2}$, the lowest Landau level approximation, which neglects all
contributions to the order parameter from the upper Landau levels and takes
into
account only the lowest one,  is justified for the strong magnetic field due to
the absence of mixing between Landau levels.   We consider this strong magnetic
field case and generalize this complex order parameter $\phi$ to a  complex
matrix $\tilde\phi_{ij}$.

The Hamiltonian is given by
\begin{equation}
\label{eq:hamil}
H(\tilde\phi)={1\over 2m} {\rm Tr} |(-i\nabla_\mu -eA_{\mu})\tilde\phi |^2 +
               \alpha {\rm Tr} |\tilde\phi|^2 +{\beta\over 2N} {\rm Tr}
               |\tilde\phi|^4
\end{equation}
where $\tilde\phi$ is a rank $N$ complex matrix and $\mu=x,y$.  We denote the
charge of  superconductor by $e$, and this charge $e$ is of course twice of the
charge of a single electron.  Here the abelian gauge field $A_\mu$ is a vector
potential of a magnetic field.  $\alpha$ and $\beta$ are usual GL parameters.
We
choose the Landau gauge ${\bf A}=(0,Bx)$.   The complex matrix $\tilde\phi$ is
written by the projection to the lowest Landau level as,
\begin{equation}
\tilde\phi_{ij}=\sum_q M_{ij}(q)(L_y )^{-1/2} \left( {eB\over \pi}
                \right)^{1/4} e^{iqy} \exp \left[ -{eB\over 2}
                (x-{q\over eB})^2 \right]
\end{equation}
where $M_{ij}$ is a $N\times N$ complex matrix and we put $m=1$ and $\hbar =1$.
Then the Hamiltonian is rewritten as
\begin{eqnarray}
\label{eq:Gauged}
H(M) &=& \alpha_B \sum_q {\rm Tr} |M(q) |^2 + \sum_{q_i} {\beta\over 2L_y N}
      \left({eB\over 2\pi }\right)^{1/2} \nonumber\\
&\times& \exp \left[ -{1\over 2eB}(\sum q_i^2 -{1\over 4}
         (\sum q_i )^2 )\right]\nonumber\\
&\times& \delta_{q_1 +q_2 ,q_3 +q_4 }{\rm Tr} \left[ M^\ast (q_1 )M(q_3 )M^\ast
(q_2 ) M(q_4 )\right]
\end{eqnarray}
where $\alpha_B = \alpha + {eB\over m}$ is related to the reduced temperature
$[T-T_c (B)]/T_c (B)$.  The free energy $F$ is obtained
\begin{equation}
F= -{1\over N^2} \ln Z
\end{equation}
\begin{equation}
Z = \int dM e^{-H} .
\end{equation}

The perturbation series for the free energy is expanded by a new variable
defined by
\begin{equation}
\label{eq:defg}
g = {\beta eB\over 4\pi\alpha_B^2}.
\end{equation}
This is the same as the expansion parameter in $N=1$ case\cite{large,ruggeri}.
Since we have generalized the GL order parameter to a matrix variable, the
perturbation series has a $N$-dependence.  In the diagrammatic language, the
planar diagrams are obtained in the large $N$ limit.   We have nonlocal
interaction in Eq.(\ref{eq:Gauged}) and have to perform Gaussian integration,
which is equivalent to  count the  number of Euler path $T$ of each
diagrams\cite{ruggeri}. Thus there appears additional factor $1/T$ to the
combinatorial factor. In the large $N$ limit, we have to select only planar
diagrams among various terms which  have been worked before by considering $N$-
dependence\cite{large}.

Here we follow the new calculational  method instead of the selecting planar
diagrams as proposed in ref.\cite{hikami1}.  Except the factor $T$ of the
number
of Euler path, the combinatorial factor of each diagram in the planar limit
becomes the same as one matrix model. There is no difference between a complex
matrix model and a Hermitian matrix model in the leading order except a trivial
factor $2$. It  has been recognized that the renormalized expansion simplifies
remarkably the diagramatic expansion for  matrix models : one needs to consider
only the irreducible diagrams to obtain the perturbation series  of the free
energy.  This renormalized expansion can be  applied successfully to our case,
since the extra  factor $1/T$ due to the Euler path is factorized. This
factorization is easily understood for any diagrams by the definition of Euler
path. Therefore, we follow the procedure of the renormalized  expansion method
\cite{hikami1}, which gives a remarkably simplified method for the
perturbation
expansion.

We introduce the equivalent $2N^2$-real vector model which is expressed by the
$2 N^2$ dimensional  real vector field ${\bf r}$ and the effective Hamiltonian
for large-$N$ limit is written as
\begin{equation}
H_{\mbox{eff}} = x + {1\over N^{4k-2}}\sum_{k=1}^\infty f_k g^k x^{2k}
\end{equation}
where
\begin{equation}
x = {\bf r}^2 = \left<{\rm Tr} M^\ast M \right>.
\end{equation}
Note that ${\rm Tr}M^\ast M$ is the sum of the square of the absolute value for
the matrix element. We integrate out the angular variables of this $2N^2$-
dimensional coordinates by keeping the radial part $|{\bf r}|$. We choose the
appropriate coefficients $f_k$ so as to make the free energy in the large $N$
limit  becomes the same as that of the original gauged matrix model defined by
Eq.(\ref{eq:Gauged}). This coefficient $f_k$ turns out to be determined
by the irreducible diagrams in the original matrix model.

By the saddle point method, the free energy for the Hamiltonian
Eq.(\ref{eq:Gauged}) becomes in the large $N$ limit,
\begin{equation}
\label{eq:n2vec}
{F\over N^2}=x +\sum_{k=1}^\infty f_k g^k x^{2k} - \ln x
\end{equation}
where we have replaced $x \rightarrow N^2 x$. The saddle point equation is
derived as
\begin{equation}
{x\over N^2}{\partial F\over \partial x} = x +\sum_{k=1}^\infty f_k (2K)g^k
x^{2k} -1 =0.
\end{equation}
This relation is also expressed simply by
\begin{equation}
\label{eq:saddleq}
x=1-2gy
\end{equation}
where we define
\begin{equation}
y={1\over N^2}{\partial F\over \partial g}.
\end{equation}

The relevant irreducible diagrams in each order of $g$ is reduced much in
number
and in Table I  we give these numbers of diagrams.  The self-energy part is
completely renormalized into the quantity $x=<{\rm Tr} M^\ast M >$ and only
irreducible diagrams should be taken into account.  In other words, our
irreducible diagrams do not contain the self-energy diagrams.   We briefly
explain how we obtain the irreducible planar diagrams : we generate $n$-th
order
planar irreducible diagrams from $(n-1)$-th order ones by cutting two lines and
add one vertex. The choice of two lines to be cut is restricted by the
condition
that generated next order  diagrams should also be planar, and  the directions
of the lines are uniquely determined also by this condition.  Firstly we make
the generated graphs undirected and drop all the isomorphic graphs obtained by
this process.  Then for the remaining graphs, we choose properly directed
graphs
and calculate the combinatorial factor by counting the number of isomorphisms
of
the graphs.  The number of Euler path of a given graph is determined  by
evaluating the determinant of the adjacency matrix\cite{ruggeri}.  In Fig.1  we
show planar irreducible diagrams to eighth order, where the direction of the
line (arrow) is not shown and the line is denoted simply by a sigle line
instead
of  double lines.

Up to the eighth order we obtain the perturbation series as
\begin{eqnarray}
\label{eq:gmatfree}
{F\over N^2}&=& -\ln x +  x + 2gx^2 -{1\over 2}g^2 x^4 +{8\over 9}g^3 x^6
-2{{3}\over 5}g^4 x^8 + 9{{149}\over {175}}g^5 x^{10}  \nonumber\\
& &  -43{{488}\over {715}}g^6 x^{12}
 +212{{48357908}\over {101846745}}g^7 x^{14} - 1144{{21940452333362}\over
{33393321606645}} g^8 x^{16}
\end{eqnarray}
where the coefficients are evaluated solely from the irreducible diagrams.

It may be instructive to see how this renormalized expansion method gives an
efficient result  for the case of one matrix model where the exact solution is
known :  We now consider the case without a magnetic field in $d=0$ dimension.
In the one matrix model, the Hamiltonian becomes with Hermitian matrix $M$
\begin{equation}
\label{eq:hamil_one}
H= {1\over 2}{\rm Tr} M^2 + {g\over N}{\rm Tr} M^4
\end{equation}
from which we obtain the perturbation series for the free energy in the large
$N$ limit with  the same irreducible diagrams.  (Here we omit the factor $1/T$
for the Euler path.) Since $M$ is Hermitian in Eq.(\ref{eq:hamil_one}), there
appears a difference of a factor $2$ compared to the complex matrix case except
the number $T$ of Euler path,
\begin{equation}
\label{eq:onemat}
{F\over N^2} = -{1\over 2} \ln x +{1\over 2} x + 2gx^2 - 2g^2 x^4 +
    {32\over 3}g^3 x^6 - 96 g^4 x^8 + \cdots
\end{equation}
The saddle point equation is given by
\begin{eqnarray}
\label{eq:onesdd}
{x\over N^2}\left( {\partial F\over \partial x}\right ) &=& -{1\over 2} +
{1\over 2}x
+4gx^2 - 8g^2 x^4 + 64 g^3 x^6 - \cdots \nonumber\\
&=& 0
\end{eqnarray}
{}From Eq.(\ref{eq:onemat}) we obtain for $y=\partial (F/N^2 )/\partial g$,
\begin{equation}
\label{eq:wai}
y= 2x^2 -4gx^4 + 32g^2 x^6 -384 g^3 x^8 + O(g^4).
\end{equation}
In the iteration for small $g$ with Eq.(\ref{eq:wai}), $x^2$ is reexpressed as
\begin{equation}
\label{eq:x2}
x^2 = {1\over 2}y + {1\over 2}gy^2 - g^2 y^3 +{9\over 2}g^3 y^4 -\cdots
\end{equation}
By the investigation of the ratio of the coefficients in the series
Eq.(\ref{eq:x2}),
we find that the ratio $R_k = c_k / c_{k-1}$ is given by $R_k = -12 + 30/k$
($k\ge 3$).
Then we have a following closed equation,
\begin{equation}
\label{eq:exact}
gx^2 = {1\over 108}\left[ (1+12gy)^{3/2} -1\right ] + {gy\over 3}.
\end{equation}
{}From this result, we can determine the coefficients of the free energy in
Eq.(\ref{eq:onemat}) up to any order of $g$ for the one matrix model.

We now come back to the case with a strong magnetic field.  Including the
factor
$1/T$ of Euler path, we are able to check the result of Eq.(\ref{eq:gmatfree})
comparing with Eq.(\ref{eq:onemat}). We repeat the same procedure as the one
matrix model. From Eq.(\ref{eq:gmatfree}) the saddle point equation becomes
\begin{eqnarray}
\label{eq:gmatsdd}
{x\over N^2}\left( {\partial F\over \partial x}\right) &=& -1 + x + 4gx^2 -2g^2
x^4 +{{16}\over 3}g^3 x^6
   -{{104}\over 5}g^4 x^8 + O(g^5 ) \nonumber\\
&=& 0
\end{eqnarray}
We rewrite this equation by the iteration for small $g$ as
\begin{equation}
x = 1 -4g + 34 g^2 -{1120\over 3}g^3 + \cdots
\end{equation}
Using the relation of Eq.(\ref{eq:saddleq}) (note $y$ is defined by $y=\partial
(F/N^2) /\partial g$), the free energy is obtained in the large $N$ limit,
\begin{eqnarray}
\label{eq:gmatfree_g}
{F(g)-F(0) \over N^2}  &=& 2g -{{17}\over 2}g^2 +62{2\over 9}g^3
                -585{14\over 15}g^4 +6396{272\over 525}g^5 \nonumber\\
     & & -77001{3702\over 5005}g^6 +993805{{7137857}\over {33948915}}g^7
         -1.351344723 \times 10^7 g^8
\end{eqnarray}
We correct here the result of the previous calculation of order $g^7$ and $g^8$
in ref.\cite{hikami1},  where small deviation was represented.  It is
remarkable
that we easily obtain the series expansion by  considering rather small
diagrams.

\sect{The critical point}

Our main object is to apply the matrix model to the investigation of the flux
lattice melting transition of a superconductor in a strong magnetic field for
two dimensions.   Hereafter we consider the gauged matrix model with
negative mass $\alpha_B <0$ and positive coupling case $g>0$ .

{}From the study of the large $N$ limit of the two dimensional $U(N)$ lattice
gauge theory  \cite{gross,hikami3} and one matrix model\cite{sima}, it is known
that there is a critical point of $g$ and the free energy shows different
behaviors in the small  coupling ($g<g_c $) and the strong coupling ($g>g_c $)
regions.   Also the third derivative of the free energy with respect to the
temperature  becomes discontinuous at this point  and third order phase
transitions are observed in these models. This phase transition is equivalent
to the freezing of the saddle point value of $x$.   In the equivalent $ 2N^2$-
vector model representation, $x$ is frozen below this critical point as
\begin{equation}
\label{eq:ansatz}
x = {s\alpha_B \over g}
\end{equation}
where $s$ is a certain negative value to be determined.

We again go back to the relation between $x=<{\rm Tr}M^\ast M>$ and $y=<{\rm
Tr}(M^\ast M)^2 >$.  From Eq.(\ref{eq:gmatfree}) $y$ is expanded by
\begin{equation}
y = {1\over N^2}{\partial F\over \partial g} = 2x^2 -gx^4 + {8\over 3}g^2 x^6
- {52\over 5}g^3 x^8  + \cdots
\end{equation}
This relation remains true even when we introduce the negative $\alpha_B$.
We reexpress this perturbation series as
\begin{eqnarray}
\label{eq:gx2}
gx^2 &=& {1\over 2}gy +{1\over 8}g^2 y^2 -{5\over 48}g^3 y^3 +{299\over 1920}
g^4 y^4 -{8249\over 26880}g^5 y^5 \nonumber\\
& & + {4624511\over 6589440}g^6 y^6 -{83706754319\over 49662412800}g^7 y^7
\nonumber\\
& & +  {{26766869658031714037}\over {5471161812032716800}} g^8 y^8 +
O(g^9 ).
\end{eqnarray}

For the discussion of the region below the mean field transition temperature,
$\alpha_B <0$,  we write the free energy with an explicit $\alpha_B$-dependence
in the term of order $x$,
\begin{equation}
F = -\ln x + \alpha_B x +  2gx^2 - {1\over 2} g^2 x^4 + \cdots
\end{equation}
In this case, we change the definition of $g$ from the previous one
Eq.(\ref{eq:defg})  to
\begin{equation}
g= {\beta eB \over 4\pi }
\end{equation}
The saddle point equation becomes,
\begin{equation}
\alpha_B x = 1 -2gy
\end{equation}
Thus we have for the temperature $\alpha_B <0$, with Eq.(\ref{eq:ansatz})
\begin{equation}
\label{eq:sadl2}
gy ={g-s\alpha_B^2 \over 2g}
\end{equation}

As shown in ref.\cite{hikami1}, the critical point is obtained exactly for the
one matrix model from the  exact equation of Eq.(\ref{eq:exact}) by insertion
of
Eq.(\ref{eq:ansatz}) and $x=1-4gy$. For our present model, the exact equation
is
unknown.   Thus we approximate Eq.(\ref{eq:gx2}), which corresponds to
Eq.(\ref{eq:exact}) for  one matrix model by the Pad\'e form, and repeat the
same analysis which determines the critical point. Approximated Pad\'e form of
Eq.(\ref{eq:gx2}) becomes
\begin{equation}
\label{eq:pade}
gx^2 = {1\over 2}gy \cdot{{1+b_1 gy + b_2 (gy)^2 + \cdots + b_{p-1} (gy)^{p-
1}}\over        {1+ c_1 gy + c_2 (gy)^2 +\cdots + c_q (gy)^q}} \qquad \mbox{:
$[p,q]$ Pad\'e}.
\end{equation}
We put Eq.(\ref{eq:sadl2}) into this equation then we have
\begin{equation}
\label{eq:plot}
-sz = {{(1+z)(B_0 + B_1 z + \cdots + B_{p-1} z^{p-1} )}\over
      {C_0 + C_1 z + \cdots + C_q z^q }}
\end{equation}
where $z=-s\alpha_B^2 /g$ and the coefficients ${B_0, B_1 , \cdots, C_0, C_1 ,
\cdots}$ are obtained by the Pad\'e coefficients in Eq.(\ref{eq:pade}). The
degenerate solution of $z$ is obtained when the line $-sz$ becomes tangent to
the  curve given by the r.h.s. of Eq.(\ref{eq:plot}) (Fig.2). This degenerate
solution gives the  critical point $z_c$. We apply various Pad\'e methods for
the r.h.s of Eq.(\ref{eq:plot}).  In Table II we list the critical point $z_c$
as a Pad\'e table form.  From this table we see that the convergency of the
Pad\'e which involve even number of terms of original series is good.  In the
case that $p\le q$, there appears a pole in the positive $z$ and we consider
the critical point is effected by this pole. We find the transition point  $z_c
=5.6$, $s_c = -0.4$ and $\alpha_B^2 /g =14$ by $[5,3]$ Pad\'e.

In the one matrix model, we have an explicit  expression which corresponds to
Eq.(\ref{eq:plot}) as \cite{hikami1}
\begin{equation}
-sz = {1\over 108}[(4+3z)^{2/3} -1] + {1\over 12} + {z\over 12}
\end{equation}
and the exact transition point $z=4$, $s=-1/4$ and $\alpha_B^2 /g =16$.
The exponent $3/2$ is of course related to the critical exponent of the free
energy, $1-\gamma_{\rm st}$, where $\gamma_{\rm st}$ is the  exponent of the
string susceptibility $\gamma_{\rm st}=-1/2$.  In our case, this string
susceptibility exponent is considered to be zero, since our system has a
central
charge $c=1$ as shown in ref.\cite{hikami1}. Thus, we expect that the
singularity of the r.h.s. of Eq.(\ref{eq:plot}) at the negative  $z_c$ becomes
$(1-z/z_c )\ln (1-z/z_c )$.  Then we must consider the logarithmic singularity
and our simple Pad\'e form lose the validity.  However, we are interested in
the region of positive $z_c$, and  the assumption of Eq.(\ref{eq:plot}) may be
valid.

In the previous paper\cite{HFL}, we discussed the flux lattice melting point
from  the direct calculation of the perturbation series for the GL free energy.
In this study we used the convenient reduced temperature $y_t$ which is defined
as
\begin{equation}
y_t  = {\alpha_B \over \sqrt{eB\beta /2\pi}}.
\end{equation}
This reduced temperature $y_t$ is expressed by the definition of $z$ and $s$,
\begin{equation}
y_t = \sqrt{z\over -2s}
\end{equation}
Table III shows the obtained critical point in term of reduced temperature
$y_t$
by the various Pad\'e method.

For $g<g_c $, the saddle point $x_c$  is frozen, and $s$ becomes a constant
$s_c$. The derivative of the free energy $y=\partial F/\partial g$ is given by
Eq.(\ref{eq:sadl2}) as
\begin{equation}
y= {1\over 2g}-{s_c \alpha_B^2 \over 2g^2 }.
\end{equation}
Then the free energy becomes for the low temperature phase as
\begin{equation}
\label{eq:lowfree}
{F(g)-F(0) \over N^2}={1\over 2}\ln g + {s_c \alpha_B^2 \over 2g}.
\end{equation}
The phase transition occurs at $z=z_c=-s_c \alpha_B^2 /g_c $.

In order to compare our model in the large $N$ limit with the usual $N=1$ case,
we write  the free energy $F$ and the specific heat $C$ in modified parameters
in  ref.\cite{large,HFL}.  Instead of $g=\beta eB/4\pi \alpha_B^2$ in
Eq.(\ref{eq:defg}),  we introduce a new variable $\tilde g= \beta eB/4\pi
\tilde\alpha^2$, where  $\tilde\alpha$ is a mass of Hartree-Fock approximation
and related to $\alpha_B$ as
\begin{equation}
\alpha_B = \tilde\alpha (1-4\tilde g)
\end{equation}
\begin{equation}
\tilde g ={\beta eB \over 4\pi\tilde\alpha^2}
\end{equation}
Then the Gibbs free energy $G$ in two dimensions is given by
\begin{eqnarray}
G & = & {eB\over 2\pi}\left[ \ln\left({\alpha_B \over \pi}\right) + f\left(
{\beta eB\over 4\pi\alpha_B^2}\right)\right] \nonumber\\
& = & {eB\over 2\pi}\left[ \ln\left({\tilde\alpha\over \pi}\right) + \tilde f
(\tilde g)\right ]
\end{eqnarray}
The function $f(g)$ is equal to $[F(g)-F(0)]/N^2 $ in Eq.(\ref{eq:gmatfree_g}).
We have changed the variable $g$ to $\tilde g$, since $g$ is divergent at
$\alpha_B  =0$.  The new $\tilde f(\tilde g)$ is readily obtained from
Eq.(\ref{eq:gmatfree_g}) as
\begin{eqnarray}
\tilde f(\tg) &=& -2\tg -{1\over 2}\tg^2 + {8\over 9}\tg^3 -4{3\over 5}\tg^4
+ 28{272\over 525}\tg^5 -188{5393\over 6435}\tg^6 \nonumber\\
& & + 1309{133391\over 2263261}\tg^7 - 9471{128589189285247\over
166966608033225}\tg^8
\end{eqnarray}
This expression should be compared to the result for $N=1$ case; here we
represent  the previous result \cite{large},
\begin{eqnarray}
\label{eq:free_pre}
 \tilde f(\tg) \bigg |_{N=1} &=& -2\tg -\tg^2 +4{2\over 9}\tg^3 -39{29\over
30}\tg^4 + 471.396594517 \tg^5 \nonumber\\
& & -6471.56257496\tg^6 + \cdots
\end{eqnarray}
It is easily seen that the $N=\infty$ case, the series is convergent, while the
series for $N=1$ in Eq.(\ref{eq:free_pre}) is an asymptotic expansion.

For the low temperature region, we have from Eq.(\ref{eq:lowfree})
\begin{eqnarray}
{G\over {\left({eB\over 2\pi}\right)}} &=& {1\over 2}\ln \left[ {\tg\over (1-
4\tg)^2}
\right] + {s_c (1-4\tg)^2\over 2\tg}\nonumber\\
&\simeq& -{1\over 2}\ln \tg+ 8\tg s_c
\end{eqnarray}
for $\tg \rightarrow \infty$.  Since we have found $s_c \simeq -0.4$, this
behavior coincides with the low temperature free energy \cite{HFL,thou}, of
$N=1$ case,
\begin{equation}
{G\over {\left( {eB\over 2\pi}\right)}} = -{1\over 2}\ln\tg -{4\tg\over 1.16}
\end{equation}
where a factor $1.16$ is the Abrikosov ratio $\beta_A =<|\psi|^4 >/<|\psi|^2
>^2$ for the triangular lattice.

The specific heat $C$, normalized by $\Delta C = v/\beta$ ($v$ is volume), is
obtained by the derivative of entropy $S$
\begin{equation}
{C\over \Delta C}={1\over \Delta C} {dS\over d\alpha_B}
\end{equation}
and the entropy $S$ is given by
\begin{equation}
S=-<|\psi|^2 > = -{\partial G\over \partial \alpha_B}
\end{equation}

It may be interesting to compare the $N=\infty$ result of the specific heat
with
the  previous result of $N=1$ case.  The reduced temperature $y_t$ defined by
Eq.(\ref{eq:wai}) becomes
\begin{equation}
y_t = {\alpha_B \over \sqrt{{eB\beta\over 2\pi}}} = {(1-4\tg)\over \sqrt{2\tg}}
\end{equation}
The specific heat in the high temperature region is analized by the Pad\'e
method, which has been studied before for $N=1$ case.  Remarkably the large $N$
result agrees with the previous result of $N=1$ case.  For the low temperature
region, the specific heat is  simply given by
\begin{equation}
\label{eq:lowc}
{C\over \Delta C} \simeq {1\over y_t^2} -2s_c
\end{equation}
In Fig. 3, we represent the curves of the specific heat in the low temperature
and in the  high temperature regions.  The values in the high temperature
region
are obtained by $[5,3]$ Pad\'e result of Eq.(\ref{eq:plot}).  In Fig. 3 , we
represent also the curve of the specific heat of the $N=1$ case.  There is a
phase transition of third order at  $y_t \simeq -2.7$ in the large $N$ limit.

\sect{Discussion}

In this paper, we have developed a new series expansion for the matrix
Ginzburg-
Landau model and in the large $N$ limit, we have obtained a phase transition
which corresponds to the superconductor transition in two dimensions.  We have
found the remarkable agreement about the specific heat between the large $N$
limit and the usual GL model ($N=1$) in the high temperature region and in the
low temperature region. The phase transition point for $N=\infty$ case is
higher
than the value of $N=1$. We have obtained the transition point $y_t = -2.7$,
while we have $y_t =-10$ for the usual Ginzburg-Landau model of $N=1$.

It is easy to evaluate the next order $1/N^2 $, which corresponds with the
diagrams of genus one.  Also it is interesting to perform the numerical
simulation by Langevin method or Monte Carlo method for the matrix Ginzburg-
Landau model and to find the melting transition point which depends upon $N$.
We will represent this simulation result elsewhere.

As a theoretical interest, our matrix model may be interesting in several
points. (i) In one matrix model, the phase transition corresponds to the
singularity of the density of state,  where the density of the eigenvalue of
the
Hermitian matrix has a gap at the band center\cite{sima}.  Our gauged model is
considered to be similar, although the eigenvalue representation is difficult
due to the complex matrix.  (ii)The matrix  model in the large $N$ limit
represents the string behavior as a random  surface in the double scaling limit
at the negative $g$.  We have discussed the phase  transition at a positive
$g$,
and also the double scaling limit is expected at the phase transition point.
In
the superconductor, the vortex is a string, and it is  interesting to think the
phase transition, which we found in the large $N$ limit, is related to the
string theory. We note that in a different context, the analogy of the phase
transition in a  strong magnetic field to string theory has been
discussed\cite{russo}. As same as other matrix models\cite{gross,hikami3}, the
transition may correspond to the condensation, like ideal Bose-Einstein gas.
It
is interesting to consider further the tachyon condensation for our phase
transition. Our result of the new renormalized expansion becomes useful for the
phase transition on a random surface in the case of the central charge $c>1$.
(iii) In the presence of the impurity, the melting phase transition is
considered to be a second order \cite{fisher,ASL}, and a vortex glass or a
gauge
glass phase appears instead of the Abrikosov vortex lattice phase.  This gauge
glass phase has no long range order. It is interesting to note that for the
matrix model in the large $N$ limit, below the phase transition point, there is
no symmetry breaking, as seen in the density of state of the eigenvalue.  The
density of state has a gap, but still symmetric for the positive and negative
eigenvalue.  This is a common behavior seen in freezing phase transitions
\cite{somp}. Thus a matrix model in the large $N$ limit may become a model of a
gauge  glass state in the two dimensional superconductor in a magnetic field,
in
which the freezing transition is essential.

\newpage

\begin{center}
\begin{large}
{\sc\bf ACKNOWLEDGEMENTS}
\end{large}
\end{center}

\vskip 1cm

This work is supported in part by the cooperative research project between CNRS
and JSPS, and by a Grant-in-Aid for Scientific Reserch by the Ministry of
Education,  Science and Culture.

\newpage

\begin{center}
\begin{large}
{\sc\bf TABLE CAPTION}
\end{large}
\end{center}

\vskip 1cm

\begin{description}
\item{TABLE I.} Number of planar irreducible diagrams relevant in the
derivation
of Eq.(\ref{eq:gmatfree}).
\item{TABLE II.} Phase transition point $z_c$ obtained by various $[p,q]$
Pad\'e method in the $\alpha_B <0$ case.
\item{TABLE III.} Reduced relative phase transition temperature which
corresponds to the critical point $z_c $ and $s_c$ obtained by $[p,q]$ Pad\'e
method.
\end{description}

\vskip 2cm

\begin{center}
\begin{large}
{\sc\bf FIGURE CAPTION}
\end{large}
\end{center}

\vskip 1cm

\begin{description}
\item{Fig. 1} Planar irreducible diagrams up to $O(g^8 )$.
\item{Fig. 2} Approximated $[5,3]$ Pad\'e form of r.h.s. of Eq.(\ref{eq:plot})
is  plotted against $z$ by the solid line. The dotted tangent line determines
the phase transition point $z_c = 5.64$ and $-s_c = 0.4$ which is the slope of
this line.
\item{Fig. 3} The specific heat scaled by $\Delta C$ against reduced
temperature
$y_t$. For the derivation of this line $\tilde f(\tg)$ is approximated by
$[5,3]$ Pad\'e form. Low temperature side is obtained from reexpressing
Eq.(\ref{eq:lowfree}) as the function of $y_t$ and evaluating its second
derivative with respect to $y_t$. The dotted line is the specific heat for the
$N=1$ case obtained in ref.\cite{HFL}.
\end{description}

\newpage

\begin{table}[h]
\begin{center}
\begin{tabular}{|l|l|l|l|l|l|l|l|l|} \hline
    &    $O(g)$&   $O(g^2 )$& $O(g^3 )$& $O(g^4 )$& $O(g^5 )$& $O(g^6 )$&
$O(g^7 )$& $O(g^8 )$ \\ \hline
\# of planar diagrams & $1$ & $1$& $1$& $2$& $3$& $9$& $22$& $61$ \\ \hline
\end{tabular}
\vskip 5mm
\caption{}
\end{center}
\end{table}

\begin{table}[h]
\begin{center}
\begin{tabular}{|c|c|c|c|} \hline
$q\backslash p$ &$3$       &$4$       &$5$              \\ \hline
$2$             &$19.6$    &$5.64$    &$4.62$          	\\ \hline
$3$             &$5.18$    &$4.12$    &$5.64$		\\ \hline
$4$             &$4.43$    &$5.19$    &          	\\ \hline
\end{tabular}
\vskip 5mm
\caption{}
\end{center}
\end{table}

\begin{table}[h]
\begin{center}
\begin{tabular}{|c|c|c|c|}
\hline
$q\backslash p$ &$3$         &$4$      &$5$   		\\ \hline
$2$             &$-5.16$     &$-2.66$  &$-2.38$         \\ \hline
$3$             &$-2.54$     &$-2.23$  &$-2.66$		\\ \hline
$4$             &$-2.32$     &$-2.54$  &          	\\ \hline
\end{tabular}
\vskip 5mm
\caption{}
\end{center}
\end{table}


\newpage
\begin{thebibliography}{99}
\bibitem{large} E. Br\'ezin , A. Fujita and S. Hikami, Phys. Rev. Lett. {\bf
65}
(1990) 1949, {\bf 65} (1990)  2921(E)
\bibitem{ruggeri} G. J. Ruggeri and D. J. Thouless, J. Phys. {\bf F6} (1976)
2063
\bibitem{HFL} S. Hikami, A. Fujita and A.I. Larkin, Phys.Rev.{\bf B44} (1991)
10400
\bibitem{thou} D.J. Thouless, Phys. Rev. Lett. {\bf 34} (1975) 946.
\bibitem{hu2} J. Hu, A. H. MacDonald, and B. D. Mckay, Phys. Rev. {\bf B49}
(1994) 15263
\bibitem{kato} Y. Kato and N. Nagaosa, Phys Rev. {\bf B47} (1993) 2932
\bibitem{hu} J. Hu and A. H. MacDonald, Phys. Rev. Lett. {\bf 71} (1993) 432
\bibitem{franz} M. Franz and S. Teitel, Phys. Rev. Lett. {\bf 73} (1994) 480
\bibitem{sasik} R. \v S\'a\v sik and D. Stroud, Phys. Rev. {\bf B49} (1994)
16074
\bibitem{tesa} Z. Te\v sanovi\'c and L. Xing, Phys. Rev.Lett. {\bf 67} (1991)
2729.
\bibitem{oneil} J. A. O'Neill and M. A. Moore, Phys. Rev. {\bf B48} (1993) 374
\bibitem{BNT} E. Br\'ezin, D.R. Nelson and A. Thiaville, Phys. Rev. {\bf B31}
(1985) 7124
\bibitem{hikami1} S. Hikami, Prog. Theor. Phys. {\bf 92} (1994) 479
\bibitem{brezin4} "{\it 2D quantum gravity and random surfaces}", Jerusalem
winter school, edited by D.J. Gross, T. Piran and S. Weinberg, 1992, World
Scientific, Singapore.
\bibitem{hikami2} V.G. Knitzhnik, A.M. Polyakov and A.B. Zamolodchikov,  Mod.
Phys. Lett. {\bf A3} (1988) 819.   E. Br\'ezin and S. Hikami, Phys. Lett. {\bf
B283} (1992) 203.  S. Hikami and E. Br\'ezin, Phys. Lett. {\bf B295} (1992)
209.
S. Hikami, Phys. Lett. {\bf B305} (1993) 327.  C.A. Baillie and D. Johnston,
Phys. Lett. {\bf B286} (1992) 44.
\bibitem{brezin2} E. Br\'ezin and A. Zee, Nucl. Phys. {\bf B40[FS]} (1993) 613.
                  E. Br\'ezin and A. Zee, Phys. Rev. {\bf E49} (1994) 2588.
 E. Br\'ezin, S. Hikami and A. Zee, a preprint (1994) (LPENS-94-35/FNS-ITP-94-/
UT-KOMABA-94-21). C.W.J. Beenakker, Nucl. Phys. {\bf B422} (1994) 515.
\bibitem{okiji} "{\it Correlation Effects in Low-Dimensional Electron
Systems}",
16th Taniguchi Symposium, edited by A. Okiji and N. Kawakami,1994, Springer-
Verlag, Berlin Heidelberg.
\bibitem{brezin1} E. Br\'ezin, C. Itzykson, G. Parisi, and J. B. Zuber, Commun.
Math. Phys. {\bf 59} (1978) 35.
\bibitem{gross} D. J. Gross and E. Witten, Phys. Rev. {\bf D21} (1980) 446.
\bibitem{hikami3} S. Hikami and T. Maskawa, Prog. Theor. Phys. {\bf 67} (1982)
1038.
\bibitem{sima} Y. Simamune, Phys. Lett. {\bf 108B} (1982) 407.
\bibitem{affl} I. Affleck and E. Br\'ezin, Nucl. Phys. {\bf B257} (1985)
451
\bibitem{HF} S. Hikami and A. Fujita, Prog. Theor. Phys. {\bf 83} (1990) 443.
    S. Hikami and A. Fujita, Phys. Rev.{\bf B41} (1990) 6379.
\bibitem{russo} J.G.Russo, Phys. Lett. {\bf B335}, 168 (1994) and references
therein.
\bibitem{fisher}D.S.Fisher, M.P.A. Fisher and D.A.Huse, Phys. Rev. {\bf B43}
(1990) 130.
\bibitem{ASL}A. Fujita, S. Hikami and A.I.Larkin, Physica {\bf C185-189}(1991)
1883.
\bibitem{somp}H. Sompolinsky, a private communication.

\end{thebibliography}
\end{document}